\documentclass[twocolumn,showpacs,preprintnumbers,amsmath,prb,amssymb,epsf,superscriptaddress]{revtex4-1}

\usepackage{graphicx}
\usepackage[dvips]{color}
\usepackage{epsfig}
\usepackage[english]{babel}

\begin{document}

\title{Effects of lateral size and material properties on the ferromagnetic resonance response of spinwave eigen-modes in magnetic devices}
\author{K. Eason}
\email[Electronic address: ]{kwaku$\_$eason@dsi.a-star.edu.sg}
\affiliation{Data Storage Institute, Agency for Science Technology and Research (A*STAR), 117608 Singapore}
\author{M. P. R. G. Sabino}
\affiliation{Data Storage Institute, Agency for Science Technology and Research (A*STAR), 117608 Singapore}
\affiliation{Electrical and Computer Engineering Department, National University of Singapore, 4 Engineering Drive 3, 117576, Singapore}
\author{M. Tran}
\affiliation{Data Storage Institute, Agency for Science Technology and Research (A*STAR), 117608 Singapore}
\author{Y. F. Liew}
\affiliation{Electrical and Computer Engineering Department, National University of Singapore, 4 Engineering Drive 3, 117576, Singapore}
\affiliation{ National Metrology Centre, Agency for Science Technology and Research (A*STAR), 1 Science Park Drive,  118221, Singapore}
\vskip 24pt
\begin{abstract}
{We analyze the effects of lateral device size and magnetic material parameters on the ferromagnetic resonance (FMR) response.  Results presented are directly relevant to widely used FMR experimental techniques for extracting magnetic parameters from thin films, the results of which are often assumed to carry over to corresponding nanometer-sized patterned devices.  We show that there can be significant variation in the FMR response with device size, and that the extent of the variation depends on the magnetic material properties.  This explains, for example, why different experiments along these lines have yielded different size-dependent trends from damping measurements. Observed trends with increasing size and different material parameters are explained through the evolution of three distinct eigen-modes, demonstrating the respective roles of demagnetization and exchange. It is also shown that there is a crossover of dominant eigen-modes in the response signal, accompanied by conjugating edge-type modes, leading to evident effects in measured linewidth and damping. Among the sizes considered, in higher saturation magnetization, we observe as much as a $40 \%$ increase in apparent damping, due solely to device size variation.}\\
\end{abstract}
\pacs{75.30.Ds, 76.50.+g, 75.75.-c, 75.78.Cd}
\maketitle
\section{Introduction}
When a ferromagnetic material experiences a static magnetic field, its ground state is additionally split by the Zeeman energy, resulting in a condition for absorption of electromagnetic power by ferromagnetic resonance (FMR) given by $\Delta E_{Z} = g \mu_B B_{eff} = \hbar \omega$; $\mu_B$ is the Bohr magneton; $B_{eff}$ is the effective magnetic induction; and $\omega$ is the frequency of the additional AC field, which is usually in the gigahertz range.~\cite{Griffiths1946, PhysRev.71.270.2, PhysRev.73.155} For more than 60 years, tapping into such a phenomenon has proven to be a powerful tool to extract information from magnetic materials, including both static (e.g. magnetic anisotropy, effective magnetization) as well as dynamic parameters (e.g. \textit{g}-factor, damping parameter $\alpha$).~\cite{Farle1998, Celinski19976, Ultrathin2005, Narkowicz2005, Meckenstock2008, Kohler1993, Harward2011} 
The damping parameter $\alpha$, which quantifies the magnetization relaxation rate, is crucial for magnetic systems driven by finite temporal-length inputs, such as a current pulse or a modulated field on a hard-disk drive (HDD).  On a HDD, $\alpha$ determines the time evolution of magnetic grains in recording media, and in Spin-Transfer Torque Magneto-Resistive Random Access Memory (STT-MRAM), it also determines the critical current to control magnetic states.  Therefore, the performance of such data storage devices relies strongly on an accurate knowledge of damping.  In FMR experiments, the concept of apparent damping is often used for the measured quantity to make the distinction between that which is measured experimentally by inferring from linewidth and resonance  measurements and the actual intrinsic damping parameter $\alpha$ in the Landau-Lifshitz-Gilbert (LLG)\cite{SpinDyn1} equation.  Apparent $\alpha$ is investigated here directly using FMR analysis.

For some common FMR techniques, in order to obtain a strong enough signal in a measurement, the magnetic sample often needs to be orders of magnitude larger laterally than the device of interest.  However, with the advent of STT-MRAM~\cite{Kawahara2012} opening the way to magnetic storage devices with characteristic dimensions less than 20nm, there is a need to understand the fidelity of measurements of dynamic parameters determined by FMR on unpatterned thin films~\cite{IEEETM:Noh,PhysRevB.79.184404}.  While recently developed techniques such as Spin-Torque FMR~\cite{Tulapurkar2005, Kubota2008, Ishibashi2010}, making use of microwave spin-currents or even spin-orbit coupling induced torques~\cite{FangD.2011}, allow extracting magnetic properties at the device level, they are not suitable for all magnetic devices (ideal if designed to flow current, and contains few magnetic layers). Other techniques, such as Frequency- or Time- Resolved MOKE or Magnetic Resonance Force Microscopy have also been used in recent years to access structures smaller than typical thin films, which, aside from facing their own challenges~\cite{MagNanoRvw}, are also not as widespread. 
Thus, some of the FMR techniques in use are incapable of resolving nanometer sized device parameters.  And, it is unclear down to what size will measurments remain valid.  The effects of lateral size on the FMR response have been investigated experimentally in previous studies. For instance, Shaw $\it{et~al.}$ studied size effects in elliptical arrays and observed no change in apparent $\alpha$ from the patterning process\cite{PhysRevB.79.184404} down to $50$nm, while a later study by Noh $\it{et~al.}$ found that $\alpha$ changes significantly with the device size.\cite{IEEETM:Noh} It is worth pointing out that these studies use different materials.  Thus, it is still an open question as to whether FMR measurements on larger size films yield results which represent, with fidelity, the device of interest, which is often much smaller.

In this work, we resolve this discrepancy through extensive 3D finite element method (FEM) simulations, investigating the effects of lateral device size and magnetic material parameters on the FMR response. We show that both results are indeed possible and demonstrate why this is the case.  In particular, we discuss resonance frequency, linewidth, and apparent $\alpha$ behavior with size variation. We also vary the magnetic material properties using the saturation magnetization, $0.6$T$\leq~\mu_0M_{s}\leq~2.4$T, as it controls significant contributions to the resonance response, both through the demagnetization field, and equally important, through the exchange length $l_{ex}=\sqrt{2A/\mu_{0}M_s^{2}}$; $A$ is the exchange stiffness; $\mu_{0}$ is the magnetic permeability in free space.  The chosen range of $\mu_{0}M_s$ covers most materials of interest from Permalloy ($\mu_{0}M_{s}\sim{1}$T) to CoFe alloys ($\mu_{0}M_s\sim{1.8-2.2}$T)\cite{MagStohr}. As will be shown, smaller $l_{ex}$ permits higher order (i.e. more nonuniform) modal contributions into the response sooner with respect to size, and is important to consider.

We first describe the FEM model used and demonstrate a validation of it, reproducing reported experimental and simulated data. Then, we discuss the resonance behavior versus device size across different materials.  We especially emphasize the behavior of devices small enough for uniform precession modes to be dominant, as small as 10nm, then study the response through the evolution of the coexisting eigen-modes when scaling up size laterally and varying magnetic material properties.  In particular, our discussion of the corresponding resonance frequency, linewidth, and $\alpha$ response is given based on three relevant spin-wave eigen-mode evolutions. Finally, we summarize our results. 

\section{FMR Model Validation}
\begin{figure}[h]
\centering
\begin{tabular}{cc}
\epsfig{file=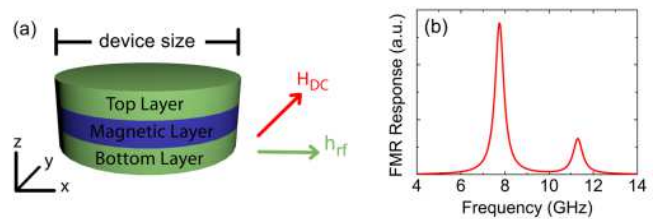,width=1.0\linewidth,clip=}
\end{tabular}
\caption{(Color online) (a) Device geometry used in our simulations: a magnetic layer (dark blue) is sandwiched between two nonmagnetic layers (light green).  For the model validation, the geometry is elliptical, with the applied field along the long axis of the ellipse, as in Ref.~\citenum{PhysRevB.79.184404}. (b) A typical FMR spectrum showing two visible resonance peaks simulated for a circular disk with diameter $\textit{d} = 150$nm, $\mu_0M_s = 1.7$T, and $H_{DC} = 0.05$T. }
\label{fig:geo}
\end{figure}

Our analysis utilizes an FEM formulation implemented within the software package SpinFlow 3D.{\cite{SF3Durl}}  The modeling procedure is as follows: first, a micromagnetic equilibrium state is computed with the applied DC field $H_{DC}$. This is followed by a computation of the spin-wave eigen-modes and associated eigen-frequencies in the vicinity of the known equilibrium state. With this information, the FMR response to a small AC field $h_{rf}$ is determined.  More details of the software formulation can be found elsewhere\cite{SF3DFMRTheory}. An important requirement of this approach is to validate the model by simulating independently reported experimental and simulated data. For this purpose, we compare our results with reported data for Permalloy (Py) nanodisks from Shaw $\it{et~al}$\cite{PhysRevB.79.184404}. The simulated device consists of a single magnetic layer sandwiched between two nonmagnetic layers, as illustrated in Fig.~\ref{fig:geo}(a). $H_{DC}$ is applied along $\vec{y}$.

\begin{figure}[h]
\centering
\begin{tabular}{cc}
\epsfig{file=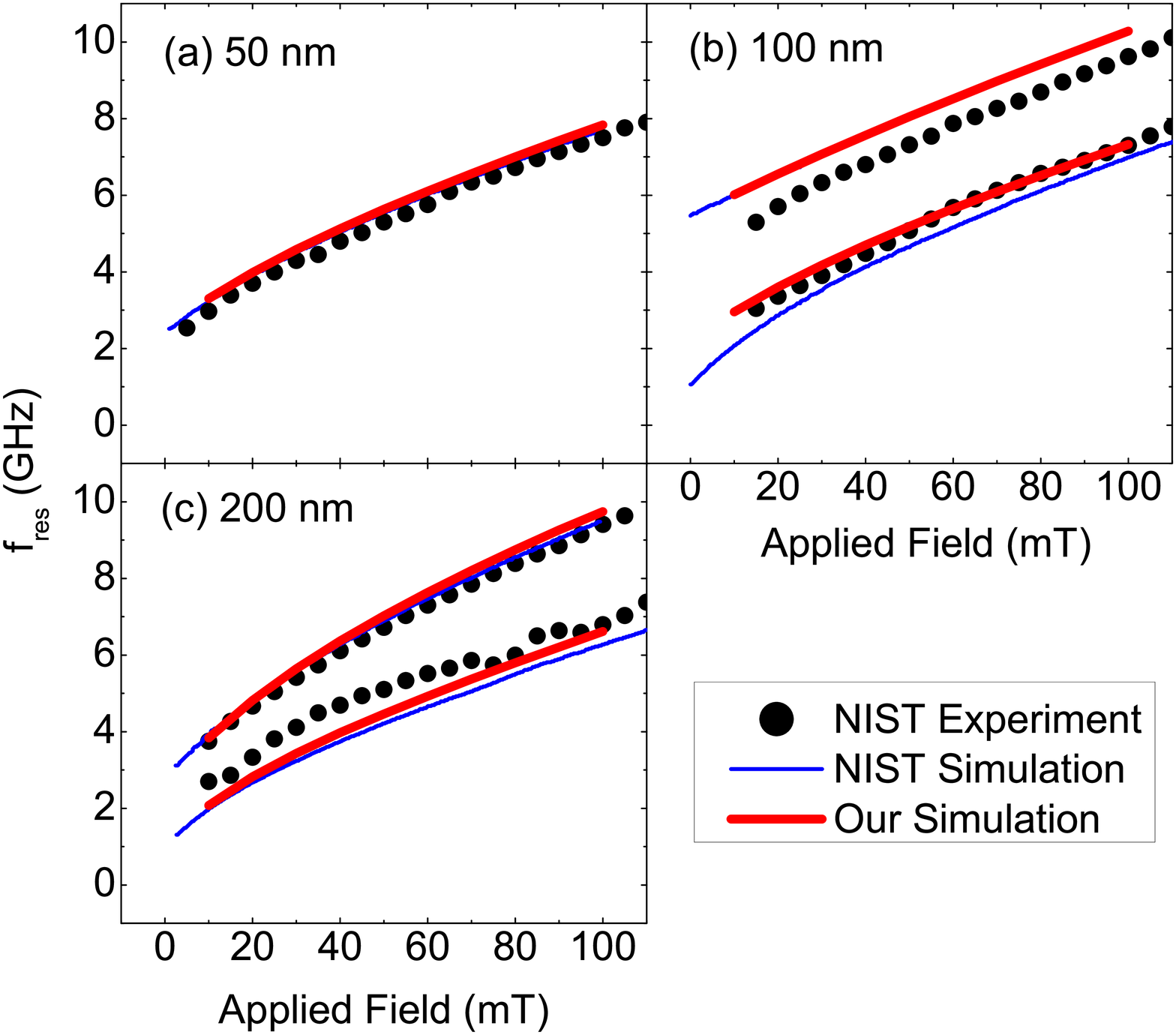,width=1.0\linewidth} 
\end{tabular}
\caption{(Color online) Comparison of simulated resonance frequency response for Py devices with nominal size (a) 50nm, (b) 100nm, and (c) 200nm as per Ref. \citenum{PhysRevB.79.184404}. Each figure includes experimental data (black circles), reported simulations (thin blue-line, NIST Simulation), and our simulation results (thick red-line, Our Simulation).}
\label{fig:val}
\end{figure}

Note that the sizes indicated are nominal sizes as defined in Ref.~\citenum{PhysRevB.79.184404}. Good agreement is achieved between our simulation results of the resonance frequencies and reported experimental data for all sizes considered (Fig. \ref{fig:val}), confirming the validity of our model. In the following sections, we present our analysis based on simulation results using the same numerical procedure to investigate the FMR response in more detail, varying lateral size and material parameters. All simulation parameters used are also summarized in Table I.   

\begin{table}[h]
\centering
\caption{Parameter values in simulations}
\label{verify_table}
\footnotesize
\begin{tabular}{l|l|l} \hline
Parameter name & Figure 2 & Other figures \\ \hline
$\mu_{0}M_s$ & $0.96$T & $0.6 - 2.4$T \\ 
exchange stiffness $A$ & $11$pJ/m & $20$pJ/m \\ 
thickness $t$ & $10$nm & $2$nm \\ 
mesh size & $5$nm & $2$nm for $\phi > 40nm$\\
					&				& $1$nm for $\phi \leq 40nm$\\
shape & elliptical & circular \\
size & a.$71\times65$nm$^{2}$ & $10 - 300$nm (diam)  \\ 
     & b.$120\times110$nm$^{2}$ & \\     
     & c.$235\times215$nm$^{2}$ & \\ 
Gilbert damping $\alpha$ & $0.01$ & $0.01$ \\ \hline
\hline
\end{tabular}
\end{table}

\section{Results and Discussion}

From here out, we discuss the FMR response of circular disk-shaped devices with varying lateral diameter $\textit{d}$, where 10nm$~\leq \textit{d} \leq~$300nm. While magnetic devices with features as small as 10nm are challenging to fabricate and access experimentally, numerical simulations avoid this difficulty and enable us to observe behavior and address whether thin film measurements obtain representative device relaxation parameters down to sufficiently small device sizes. Also note that in our FEM model, we do not take into consideration spin pumping nor inhomogeneous properties that would contribute to field linewidth broadening. Thus, our results reveal contributions due solely to size variation.

\begin{figure}[h]
\centering
\begin{tabular}{cc}
\epsfig{file=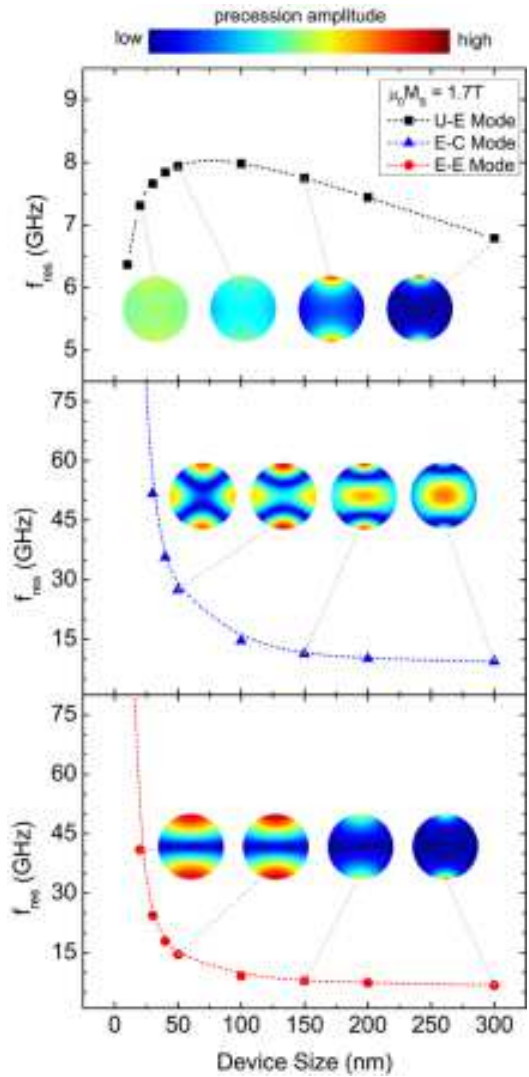,width=0.8\linewidth,clip=} \\ 
\end{tabular}
\caption{(Color online) Frequency response for the (a) Uniform-to-Edge (U-E), (b) Edge-to-Center (E-C) and (c) Edge-to-Edge (E-E) eigen-modes versus device size for $\mu_{0}M_s = 1.7$T, applied field $H_{DC}$ = 0.05T. Spatial mode profiles of precession amplitude are shown as insets.}
\label{fig:allmodeevol}
\end{figure}

A typical frequency-swept FMR spectrum $\tilde{\chi}(f)$ is shown in Fig. \ref{fig:geo}(b). There are, in general, two observable peaks with significant amplitude in the frequency range considered, indicating the presence of more than one eigen-mode. We consider three relevant eigen-modes and the resulting interplay and evolution affecting both observed peaks.  Each resonance frequency $f_{res}$ is obtained by fitting each peak of the simulated FMR response to Lorentzians of the form 
\begin{equation}
\tilde{\chi}(f)=\frac{2P}{\pi}\frac{\Delta f}{4(f+f_{res})^2+ \Delta f^2} 
\label{eq:lorentzian}
\end{equation}
where $P$ is the peak amplitude, $f$ is the frequency of the applied AC magnetic field $h_{rf}$, and $\Delta f$ is the full width at half maximum. In Fig. \ref{fig:allmodeevol}, we plot $f_{res}$ versus disk diameter for three distinct eigen-modes which contribute to the response within the frequency range considered.  Note that $\mu_{0}M_s = 1.7T$, under an applied DC field of 0.05T, enough to saturate the magnetization along $\vec{y}$. In all illustrations of mode profiles (shown as insets in Fig. \ref{fig:allmodeevol}), the color scale corresponds to the amplitude of precession - from largest (red) to smallest (blue). From the spatial distribution of precession amplitude, we identify uniform (U), edge (E) and center (C) modes, accordingly. We observe that eigen-mode profiles generally evolve as the diameter is increased, and are labelled, for instance, U-E to represent evolution with size from uniform to edge mode. Next, we discuss the origin of the evolution of modes as they are intimately tied to the behavior of $f_{res}$. 

\subsection{Evolution of Modes}

\begin{figure}[h]
\centering
\begin{tabular}{cc}
\epsfig{file=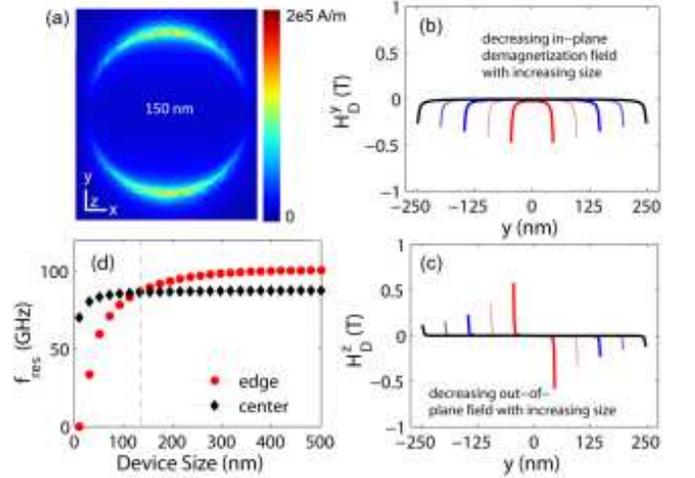,width=1.0\linewidth,clip=} \\ 
\end{tabular}
\caption{(Color online) (a)Demagnetization field (amplitude) distributions in the middle plane (along $\vec{z}$) of the magnetic layer, computed from the magnetization equilibrium distribution for $\mu_{0}M_s=1.7$T, for a 150nm device. Color scale shows $\|\vec{H_D}\|$. Computed (b) in-plane  and (c) out-of-plane demagnetization field component along $\vec{y}$ for x=0 (center line) for different sizes. (d) Resulting local ferromagnetic resonance using Eq.~(\ref{eq:Kittel}). At both center and edge, the resonance frequency always increases with increasing size.}
\label{fig:dm}
\end{figure}

For devices with sizes on the order of $l_{ex}$, which we refer to as "small" devices from hereon, the lowest energy eigen-mode is the uniform precession mode\cite{SF3DFMRTheory}. Due to the spatial distribution of the demagnetization field, the uniform-mode (U-mode) is seen to evolve into an edge-mode (E-mode) that can be observed as device size is increased. We refer to this evolution as the U-E mode.  To see the correlation more clearly, we also show the computed demagnetization field distribution in Fig.~\ref{fig:dm}(a), corresponding to the magnetization equilibrium state for a 150nm device. This distribution is typical for all sizes. Notice the inhomogeneity of the demagnetization field along the edges in the direction of $H_{DC}$, which is known to create a potential well for spin-waves tending to localize them \cite{PhysRevLett.88.047204}. The modal evolution shown in Fig. \ref{fig:allmodeevol}(a) is a direct result of the demagnetization energy as well as the distribution.  To further illustrate demagnetization edge localization, we also plot in Fig.~\ref{fig:dm}(b) and (c), respectively, computed demagnetizing field components $H_D^y$ parallel to the DC field ($\vec{y}$) and $H_D^z$ perpendicular to the film plane ($\vec{z}$) for different lateral sizes. The demagnetizing fields are clearly strongest at the edges, and comparatively weak in the interior region.  Note, however, these observations are computed for the case of nearly uniform magnetization along ($\vec{y}$).  As size is increased sufficiently beyond $l_{ex}$, nonuniform affects manifest more strongly, leading to more complex behavior. 

A second eigen-mode evolution is shown in Fig. \ref{fig:allmodeevol}(b), which evolves from an edge-mode into an eigen-mode with dominant precession amplitude within the center of the device.  This higher-order mode starts with edge precession confinement along both $\vec{x}$ (to lesser extent) and $\vec{y}$ $\sim$ a four-fold symmetry. We refer to this distinct modal evolution as the E-C mode (edge-to-center).  This particular mode shows the greatest nonuniformity, and consequently involves more exchange energy in addition to the demagnization field.  For a magnet with nonmagnetic interfacial boundaries, exchange energy is known to be lowest at the edges, due to the free-spin boundary condition.~\cite{MagBC}  In this case, the system develops more interior motion for access to more exchange energy useful to lower its total magnetic energy arising from demagnetization field effects.

The third relevant eigen-mode is shown in Fig. \ref{fig:allmodeevol}(c), in which edge-dominated precession solely persists for the entire frequency range considered.  Therefore, it begins and ends as an E-mode, also localized along $\vec{y}$ edges.  We refer to this as the E-E mode (edge-to-edge), and its evolutionary behavior is also explained by the demagnetization field, as discussed above.  Next, we analyse the implications of the modal evolutions on $f_{res}$.    

\subsection{Resonance Frequency Behavior}

To understand the resonance frequency response for non-uniform higher order eigen-modes, we consider a generalized form of the Herring-Kittel formula\cite{SpinDyn1} acounting for quantization in all three dimensions:
\begin{equation}
f_{res} = \frac{\gamma}{2\pi}[(H_{DC} + \frac{2Aq^{2}}{M_s} )(H_{DC} + \frac{2Aq^{2}}{M_s} + 4\pi{M_sF_{n,p}})]^{1/2}  
\label{eq:HKeqn}
\end{equation}
where $q$ is the wave vector given here by
\begin{equation}
q^{2} = (\frac{m\pi}{d})^{2} + (\frac{n\pi}{d})^{2} + (\frac{p\pi}{t})^{2}
\end{equation}
The parameters $m$, $n$, and $p$ are quantization numbers arising from the finite size of the device; $t$ is the device thickness; $d$ is the lateral dimension (i.e. diameter); and $F_{n,p}$ contains nontrivial demagnetization field information, which also depends on the wave vector. Note that although no simple quantization scheme can be used for structures confined in all three dimensions\cite{SpinDyn1}, the presented form of the wave vector does allow us to qualitatively understand the behavior of $f_{res}$ for higher order eigen-modes. 

In the size regime where the U-E mode is mostly uniform, the increase in $f_{res}$ is understood by considering the decreasing (with increasing size) demagnetizing field.  This decreasing trend can be seen in Fig.~\ref{fig:dm}(b) and (c).  To understand why the field decreases as size increases, consider the exact solution for the demagnization field in the case of uniform magnetization, given by\cite{SpinDyn1}   
\begin{equation}
H_{D}(\bf{x}) = -\frac{1}{4\pi}\int{\bf{M}\cdot\bf{n}\frac{\bf{x'}-\bf{x}}{\vert\bf{x'}-\bf{x}\vert^{3}}dS'}  
\label{eq:MagnInt}
\end{equation}
Consider a rectangular film with uniform magnetization $\bf{M}$ in-plane along $\vec{y}$.  In this case, only two of the six rectangular boundary faces contribute to the integral in~(\ref{eq:MagnInt}), i.e. the two faces with normal vectors $\bf{n}$ parallel to $\bf{M}$. In the center, the value is identically $0$ due to symmetry, however, away from the center and only near the edges, nonzero values exist.  Evaluating the field leads to terms of the following form    
\begin{equation}
H_{D} \propto  {M_{s}}{\frac{1}{d^{2}}dS'}  
\end{equation}
By keeping the film thickness $t$ constant, $dS'$ scales only with $d$ (instead of ${d}\times{t}$) and the result is that the demagnetization field scales as $H_{D} \propto 1/d $.  Thus, the demagnetization field decreases with increasing lateral size.  The resonance frequency consequently increases because $f_{res} \propto H_{eff} = H_{dc}-|H_{D}|$ where both predominantly lie along $\vec{y}$.  This accounts for the U-E mode increasing resonance frequency behavior in small device sizes.  However, as size increases beyond $\sim50$nm, the nonuniform mode behavior contributes more significantly. 

\begin{figure}[h]
\centering
\begin{tabular}{cc}
\epsfig{file=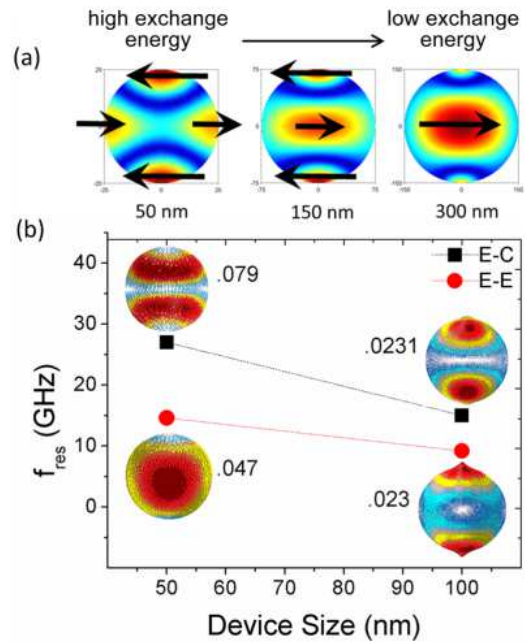,width=0.80\linewidth,clip=}
\end{tabular}
\caption{(Color online) (a) Illustration of reduction in exchange energy in high frequency E-C mode, with increasing size (50, 150, and 300nm shown) and (b) Computed gradients of in-plane normal modal component for 50nm and 100nm. Maximum gradient (in arbitrary units) shown beside the respective gradient distribution.}
\label{fig:exch}
\end{figure}

In this case, equation~(\ref{eq:HKeqn}) suggests exchange energy also contributes to $f_{res}$ in higher-order modes, through the wave-vector $q$.  In particular, it follows from~(\ref{eq:HKeqn}) that resonance frequencies generally decrease with increasing device size, as ${1/d^{2}}$.  This is in qualitative agreement with trends shown in Figures~\ref{fig:allmodeevol}(b) and (c) corresponding to higher order eigen-modes. This also applies to Figures~\ref{fig:allmodeevol}(a), after the transition to an edge mode has taken place.  In the nonuniform modes, the gradient of the eigen-modes also contains exchange information as they evolve with size.  The modes indicate precession amplitude, propertional to normal components of the magnetization (normal to $\vec{y}$).  The total exchange energy given by
\begin{equation}
E_{A} = \frac{A}{{M_{s}^{2}}}\int{({\nabla}{\bf{M})^{2}} dV} 
\label{eq:exchint} 
\end{equation}
By observing the trend in the gradient of the mode, the trend of the exchange energy can also be seen.  Figure~\ref{fig:exch}(a) illustrates the qualitative trend of the exchange associated with the E-C mode (the most nonuniform mode), indicating on the eigen-mode vector directions (with black arrows).  As size increases, the complexity of the mode reduces, and this generally reduces exchange energy density.  This behavior is equivalent to reducing the quantization numbers $m$ and $n$.  For two of the device sizes in the region of sharp decrease in $f_{res}$, the frequencies are plotted in Figure~\ref{fig:exch}(b), along with the gradient distributions of the in-plane normal mode component (x), and its maximum computed gradient indicated (in arbitray units).  This demonstrates an overall decrease in the exchange energy through the E-C mode evolution, and this contributes to a reduction in $f_{res}$ for higher order modes. 
  
\subsection{Cross-over and Conjugation of Modes}
\begin{figure}[h]
\centering
\begin{tabular}{cc}
\epsfig{file=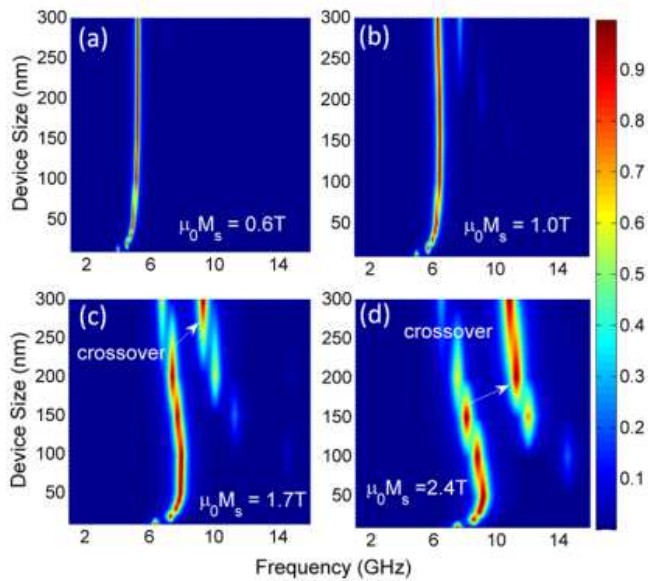,width=1.0\linewidth,clip=} 
\end{tabular}
\caption{(Color online) (a) Simulated FMR response versus device size for $\mu_{0}M_s=0.6$T, (b) 1.0T, (c) 1.7T and (d) 2.4T. White arrow indicates the location of the change in dominant modes, crossing over from U-E mode to E-C mode with increasing size.}
\label{fig:contours}
\end{figure}

Another important characteristic of the FMR spectrum, aside from the resonance frequency trend with size, is the relative amplitudes of the observed modes. The amplitude of an FMR peak is proportional to the volume of the device participating in the precession. For example, in the U-E mode, the edge precession is increasingly confined to the edges with increasing size (as can be seen in the insets of Fig.~\ref{fig:allmodeevol}) and the relative amplitude of the peak decreases, with additional contributions from a reducing demagnetization field. In contrast, the amplitude of the E-C mode increases with size relative to the other modes. Therefore, in smaller sizes, the U-E mode tends to dominate the signal, however, as size increases, the U-C mode begins to dominate.  This leads to a cross-over for the dominant mode of the FMR response signal as size increases.  The cross-over is illustrated in Fig. \ref{fig:contours}, seen clearly for $\mu_{0}M_s = 1.7$T and $2.4$T.

\begin{figure}[h]
\centering
\begin{tabular}{cc}
\epsfig{file=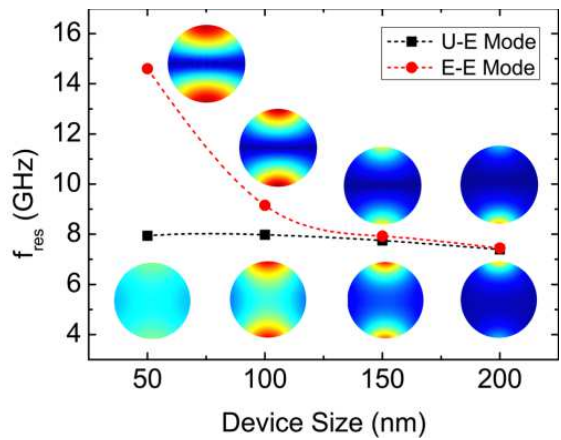,width=0.85\linewidth,clip=}
\end{tabular}
\caption{(Color online) Resonance frequency versus size for the U-E and E-E modes for $\mu_0M_s = 1.7$T, zooming in to the point of conjugation. Note the increasingly similar mode profiles of the U-E and E-E modes.}
\label{fig:conjug}
\end{figure}

Near the cross-over size, there also occurs a conjugation or joining of the U-E and E-E eigen-modes, as shown in Fig. \ref{fig:conjug}, where $\mu_{0}M_s = 1.7$T. At the point of conjugation, the mode profiles of the U-E and E-E modes become similar or quasi-degenerate and resonate at similar frequencies. Both the cross-over and the conjugation of modes can have effects on the measured linewidth as will be discussed in a later section.    
\subsection{Effect of Different $M_s$}

\begin{figure}[h]
\centering
\begin{tabular}{cc}
\epsfig{file=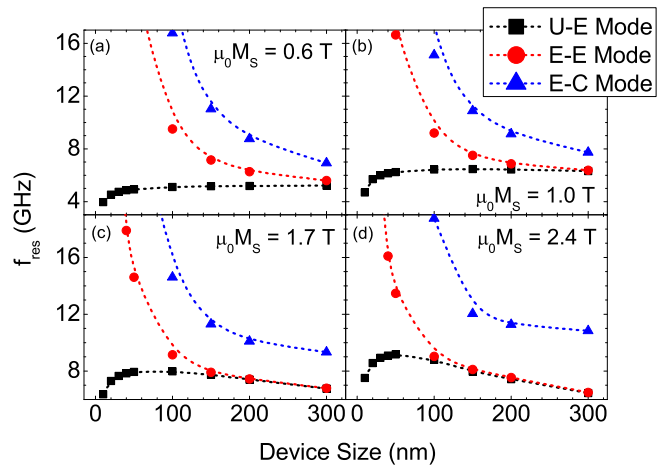,width=1.0\linewidth,clip=} \\
\end{tabular}
\caption{(Color online) (top) Frequency response of three observed eigen-modes for (a) $\mu_{0}M_s = 0.6$T, (b) 1.0T, (c) 1.7T and (d) 2.4T.}
\label{fig:fmrallmodes}
\end{figure}

One key observation in this study is that the FMR response can be sensitive to magnetic material parameters. For example, the effect of different $M_s$ on $f_{res}$ can be seen in Fig.~\ref{fig:fmrallmodes}. For the U-E mode, the device size at which an initially increasing $f_{res}$ transitions to decreasing $f_{res}$ (due to the onset of the edge mode) is smaller in materials with larger $M_s$. For the smallest $\mu_0M_s$ (0.6T), a decrease in $f_{res}$ is not observed at all in our range of sizes.  However, devices with higher $M_s$ have shorter $l_{ex}$, and it follows that non-uniform modes may contribute more significantly in smaller lateral sizes. Hence, the effect of exchange, which tends to lower $f_{res}$ as discussed above, appears earlier. Comcomitent with this, we also observe the occurence of both the modal conjugation and dominant mode cross-over at smaller sizes as $M_s$ is increased. The effects on the cross-over is evident in Fig. \ref{fig:contours}(c) and (d).

\subsection{Linewidth and Damping Behavior}

\begin{figure}[h]
\centering
\begin{tabular}{cc}
\epsfig{file=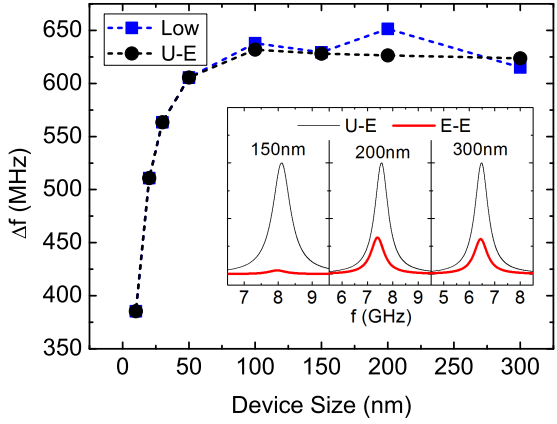,width=0.85\linewidth,clip=}
\end{tabular}
\caption{(Color online) Linewidth versus size for the low frequency peak (blue squares: with contributions from both U-E and E-E modes) compared to the linewidth of only the U-E mode (black circles). Broadening from conjugation is seen at 200nm. Inset shows the FMR spectrum for U-E (thin black) and E-E (thick red) modes in the size range where broadening may occur.}
\label{fig:conjuglw}
\end{figure}

The linewidth $\Delta f$ for each size is obtained by taking the full width at half maximum of each Lorentzian peak. For a multi-peak spectrum, we can identify separate linewidths for each peak. We typically see two significant peaks which we identify as "low" and "high" frequency peaks. The linewidth versus size behavior for the high frequency peak is mainly determined by the E-C mode. The linewidth for the low frequency peak can be more complex because both U-E and E-E modes can contribute to the signal due to their conjugation. However, below the conjugation point, both the resonance frequencies and relative amplitudes of the U-E and E-E modes are very dissimilar, and the low frequency peak linewidth is controlled mainly by U-E. As size increases, the relative amplitude of the U-E mode decreases and becomes comparable to the amplitude of the E-E mode. In some parts of this regime, their resonance frequencies do not coincide exactly. When the U-E and E-E $f_{res}$ modes do not exactly match, but are close enough, there is some artifical broadening of the measured linewidth. Once their resonance frequencies coincide, this artificial broadening is lost. To illustrate this, we plot the linewidth of the low frequency peak (which includes contributions from both U-E and E-E modes,  corresponding more closely to a measured signal) and the linewidth of only the U-E mode in Fig.~\ref{fig:conjuglw}. The insets show spectra of the U-E and E-E modes in the size range where the transient broadening may occur. Note the increased linewidth of the low frequency peak at 200nm which can be attributed to modal conjugation. 

\begin{figure}[h]
\centering
\begin{tabular}{cc}
\epsfig{file=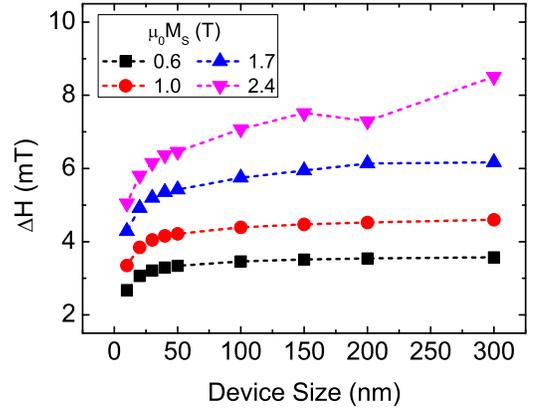,width=0.8\linewidth,clip=}
\end{tabular}
\caption{(Color online) Linewidths from FMR response versus device size for different $M_S$.}
\label{fig:linewidth}
\end{figure}

For experiments, however, looking at the dominant peak (i.e. with the highest intensity) may be more relevant because limitations in detection sensitivity may not be enough to resolve peaks other than the dominant one. Without the aid of rigorous calculation or simulation, it could be difficult to trace the origin of a particular peak in the signal; our results show that the dominant peak may be a different mode for different sizes. In the following analysis of linewidth behavior versus size for the dominant peak, we use the field linewidth $\Delta H$, which is obtained from $\Delta f$ via
\begin{equation}
\Delta H = \frac{\Delta f}{(\delta f/\delta H)}
\label{eq:linewidthconv} 
\end{equation}
The denominator is calculated from the general Kittel equation\cite{PhysRev.73.155}
\begin{equation}
f = \sqrt{(H + H_1)(H + H_2)}
\label{eq:Kittel} 
\end{equation}
where $H$ is the applied DC field and $H_1$ and $H_2$ are stiffness fields. 
We plot the linewidth of the dominant peak for different $M_s$ values in Fig.~\ref{fig:linewidth}.  We observe a clear increasing trend, with slightly more complex behavior for $\mu_0M_s = 2.4$T. Because the linewidth is given by
\begin{equation}
{\Delta}H = {\Delta}H_{0} + \frac{4\pi{\alpha}f_{res}}{\gamma}
\label{eq:delH}
\end{equation}
and is therefore proportional to $f_{res}$, we find that the change in dominant mode, which is accompanied by a distinct jump to higher $f_{res}$, results in an abrupt increase in ${\Delta}H$ at around 200nm. Moreover, after the cross-over, if the linewidth follows $f_{res}$, the now-dominant E-C mode resonance frequency decreases, so the linewidth should decrease according to Eq.~\ref{eq:delH}, where ${\Delta}H_0$ is the inhomogeneous linewidth broadening due to inhomogeneities in the device parameters, e.g. dead layers and edge damage, which are not considered here. However, linewidth does not decrease with device size, as does $f_{res}$. This indicates that $\alpha$ is non-constant and is contributing to the observed linewidth behavior.

\begin{figure}[h]
\centering
\begin{tabular}{cc}
\epsfig{file=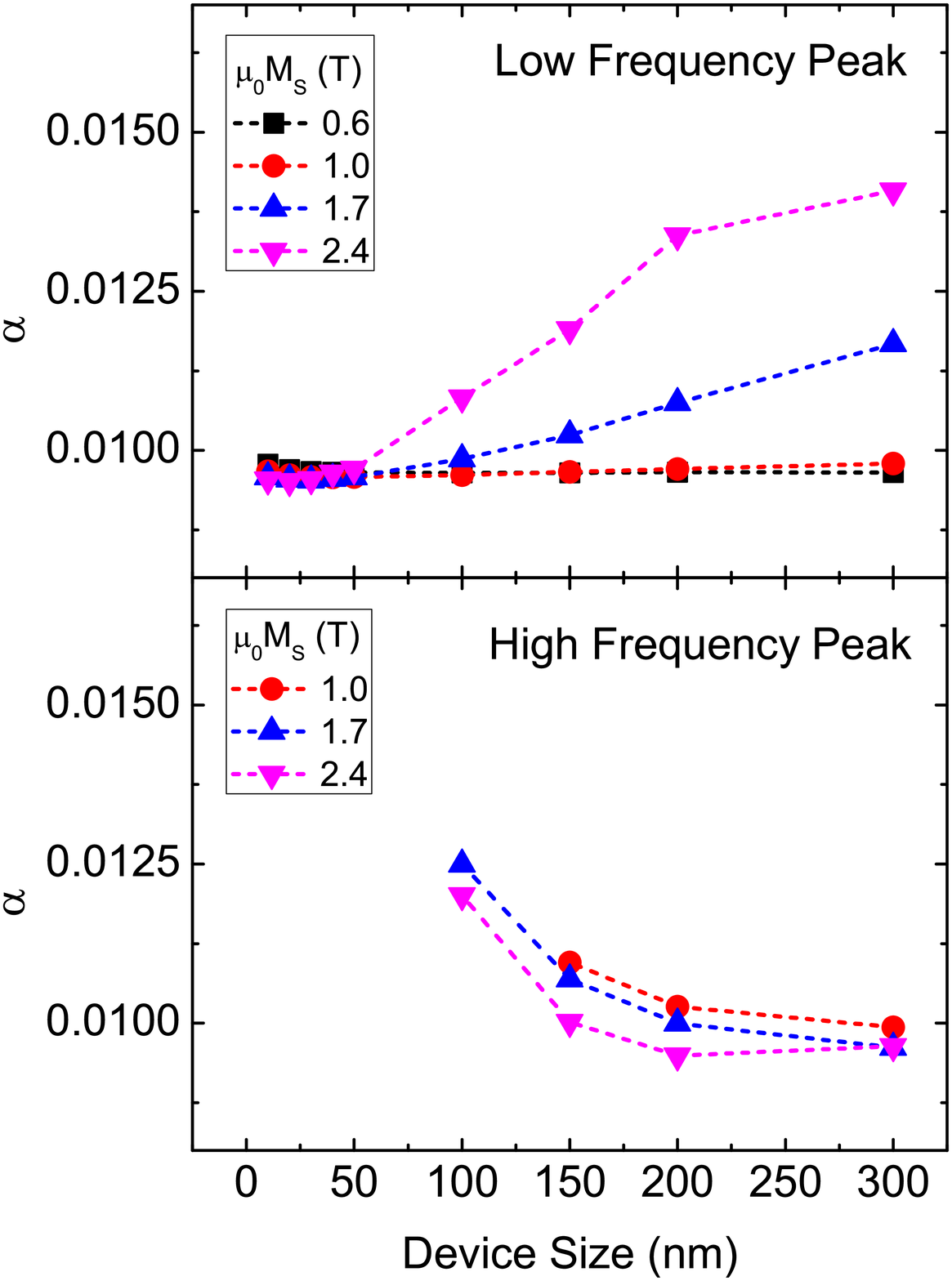,width=0.8\linewidth,clip=}
\end{tabular}
\caption{(Color online) Apparent $\alpha$ from FMR response for different $M_s$. Dashed line in (a) for $\mu_0M_s = 2.4T$ is a guide to the eye, highlighting the increased non-linearity around 200nm.}
\label{fig:alpha}
\end{figure}

We therefore extract the apparent damping $\alpha$ for each peak - by fitting applied field-dependent linewidth data to Eq. \ref{eq:delH} - to eliminate contributions from $f_{res}$. The results are shown in Fig.~\ref{fig:alpha} for different $M_s$.  For the low frequency peak, the increase in $\alpha$ is more significant in larger $M_s$. For the high frequency peak, $\alpha$ decreases monotonically for all $M_s$, approaching $\alpha = 0.01$. 

To further explain , we consider that each differential volume in the device experiences a slightly different $H_{eff}$ and therefore has a slightly different $f_{res}$ compared to its neighbor. This difference is more pronounced as $M_s$ increases because shorter $l_{ex}$ allows more non-uniformity and the effect of the demagnetization field is also stronger. The collective response - the sum of these individual responses which do not necessarily overlap - leads to broading. The low frequecy peak (mainly from the U-E mode) goes from uniform to non-uniform, leading to more broadening with size. However, for the the higher frequency peak (E-C mode), which becomes more uniform with increasing size, broadening generally decreases.  

The behavior of the linewidth for $\mu_0M_s = 2.4$T in larger sizes is actually explained by the corresponding decrease in $\alpha$ at 200nm as dominance shifts from the low frequency peak to the high frequency peak.  This also contributes to the subsequent increase at 300nm, as can be seen in Fig.~\ref{fig:linewidth}.

The effects of size and material properties can also be seen clearly in Fig.~\ref{fig:alpha}. In the range considered, materials with low $M_s$ generally show neglible variation in damping, while for higher values of $M_s$, there is a clear variation with size, indicating that the extent of the variation is sensitive to material properties, which in our case is $M_s$. In the present case, the characteristic length that controls the FMR behavior with size is $l_{ex}$. Thus, our results suggest that the different findings on $\alpha$ versus size behavior~\cite{IEEETM:Noh,PhysRevB.79.184404} are not inconsistent, but stem primarily from the use of different materials.  Our results are consistent, quanlitatively, with both reports, where Py ($\mu_0M_s = 1.0T$) shows little variation, however CoFe ($\mu_0M_s = 1.7T$) does shows a visible increase with size.

Therefore, since apparent $\alpha$ may be non-constant with size, care must be taken in using film measurements for device-related parameters. For a multi-peak FMR response, $\alpha$ of the dominant peaks in the extremes of the size spectrum considered here seem to converge to the same value, however, they do so on opposite ends of the size range. Take for example, 40nm where the low frequency peak is dominant, and 300nm where high frequency peak is dominant. Both values approach $\alpha = 0.01$ for all $M_s$ values considered. Although the applicability of film measurements to the device level seems possible, this, of course, assumes proper accounting of all other contributions to the linewidth. So, there can also be close agreement between damping measurements from film and from devices below a certain size in some materials.\cite{sun:07C711}

\section{Conclusion}
We have investigated the effects of device size and magnetic material parameters on the FMR response using a powerful 3D FEM tool. We identified three distinct eigen-modes, the profiles of which evolve versus size and consequently lead to evident effects on the FMR response. Two of the three modes (U-E and E-E) were shown to conjoin at a given size and concomitant with this conjugation, a cross-over in terms of mode dominance is also seen. These phenomena may affect measured linewidth and damping. The amount of sensitivity to size was shown to increase as $M_s$ is increased and may be the reason behind the discrepancy from different investigations, which used different materials.  Within the size range of 10nm to 300nm, we have observed as much as a 40\% increase in apparent $\alpha$, due purely to size effects. Our results suggests that an FMR measurement can indeed lead to inflation of the intrinsic damping parameter, depending on the material used.   

\bibliography{ourbib}
\end{document}